\documentclass[]{aa}

\usepackage{graphicx}
\usepackage{hyperref}
\usepackage[varg]{txfonts}
\usepackage{caption}
\usepackage{subcaption} 
\usepackage{multicol}
\usepackage{booktabs}

\usepackage{placeins}

\usepackage{mathtools} 
\DeclarePairedDelimiterXPP\BigOSI[2]
  {\mathcal{O}}{(}{)}{}
  {\SI{#1}{#2}}

\usepackage{float}
\usepackage{amsmath}
\usepackage{xcolor}

\usepackage{pifont} 

\usepackage{lipsum}

\hypersetup{
  colorlinks=true,  
  urlcolor=blue,     
  linkcolor=blue,    
  citecolor=blue
}

\begin{document}

\title{Transient star B/R ratio and star formation history in $z\gtrsim 1$ lensed galaxies}

\author{Sung Kei Li    \thanks{keihk98@connect.hku.hk}
          \inst{1}       
          \and
          Jose M. Palencia\inst{2}
          \and
          Jose M. Diego\inst{2}
          \and
          Jeremy Lim\inst{1}
          \and
          Patrick L. Kelly\inst{3}
          \and
          Ashish K. Meena\inst{4}
          \and
          James Nianias\inst{1}
          \and
          Hayley Williams\inst{3}
          \and
          Liliya L.R. Williams \inst{3}
          \and
          Adi Zitrin \inst{4}
          \and 
          Thomas J. Broadhurst \inst{5, 6, 7}
          }
   \institute{   Department of Physics, The University of Hong Kong, Pokfulam Road, Hong Kong
   \and
   Instituto de Física de Cantabria (CSIC-UC), Avda. Los Castros s/n, 39005 Santander, Spain
   \and Minnesota Institute for Astrophysics, University of Minnesota, 116 Church St. SE, Minneapolis, MN 55455, USA
   \and Department of Physics, Ben-Gurion University of the Negev, PO Box 653, Be’er-Sheva 8410501, Israel
   \and Department of Theoretical Physics, University of Basque Country UPV/EHU, Bilbao, Spain
   \and Ikerbasque, Basque Foundation for Science, Bilbao, Spain
   \and Donostia International Physics Center, Paseo Manuel de Lardizabal, 4, San Sebasti\'an, 20018, Spain
     }

\abstract{
The extreme magnification from galaxy clusters and microlenses therein allows the detection of individual, luminous stars in lensed galaxies as transient events, and hence provides a valuable window into the high mass stellar population in $z>1$ galaxies. As these bright stars can only be formed at specific ages, the relative abundance of transient events at blue (B) and red (R) optical wavelengths ($B/R$ ratio) can provide insights into the recent star formation history of galaxies that are not well constrained by their spectral energy distributions (SEDs). Here, we forward model the transient detection rates in an idealized mock scenario to find that the $B/R$ ratio of strongly lensed $z>1$ galaxies decreases quickly with increasing age.  This ratio has moderate sensitivity to metallicity and comparatively low sensitivity to dust attenuation, with no significant dependency on the stellar initial mass function.  Fitting model stellar populations to either the SED or $B/R$ ratio alone of ``Warhol'' arc ($z = 0.94$), we find that neither a simple single starburst nor a more complex star formation can simultaneously reproduce both constraints.  We then demonstrate that a best-fit model constrained by both the B/R ratio and SED requires a star-formation rate that has varied quite dramatically over the past $\sim$50 Myr, for which the total stellar mass formed over this time is a factor of 10 (with $2-3\sigma$ significance) different from the best-fit models to the SED alone.
Our work shows that the transient $B/R$ ratio can be used as an additional powerful constraint on the recent star formation history of higher-redshift galaxies in future works that are strongly lensed by galaxy clusters.
}


\maketitle

\section{Introduction}

Individual stars in lensed galaxies at $z \gtrsim 1$ are now being routinely discovered in deep, repeated \textit{Hubble Space Telescope} (\textit{HST}) and \textit{James Webb Space Telescope} (\textit{JWST}) imaging of galaxy clusters \citep[e.g., ][]{Kelly_2018_Icarus, Chen_2019_Warhol, Kelly_2022_Flashlights, Yan_2023, Meena_2023_Flashlights, Fudamoto_2025}. These lensed stars must be intrinsically luminous, and be subject to extreme magnifications --- from a combination of cluster macrolensing and stellar microlensing --- to overcome the cosmological distance modulus and be detected as transient events \citep{Miralda-Escude_1991, Oguri_2018, Diego_2019_extrememagnification}. %Therefore, only the brightest stars among the lensed stellar populations can be detected. 
The two primary populations that constitute the lensed stars are: (i) the blue supergiants (BSGs, effective temperature of $10,000 - 30,000\,K$), which can be seen in blue filters that probe the restframe UV/optical \citep[e.g., ][]{Kelly_2018_Icarus}; and (ii) red supergiants (RSGs, effective temperature of $3000-4000\,K$) which can be seen in red filters that cover the restframe NIR \citep[][]{Diego_2023_Elgordo}. Bright O/B stars, or other classes of evolved stars such as asymptotic giant branches and Cepheid variables \citep{Diego_2024_Cepheid} can also be detected as lensed transients, albeit much less frequently, given their lower luminosity.

Massive stars are very short-lived and so are only present in stellar populations at specific ages. For example, BSGs can only exist in stellar populations having ages $\lesssim 10\,$Myr. By contrast, RSGs are evolved stars that take time ($\gtrsim 10\,$Myr) to evolve from their progenitors. The relative abundance of massive BSGs and RSGs has been well explored in nearby galaxies (where resolved photometry can be carried out) to characterize properties such as the age and metallicity of the stellar populations therein \citep{Eggenberger_2002, Dohm-Palmer_2002}. As lensing is achromatic and does not affect the color of lensed stars\footnote{except for the case that these lensed stars might have different radii thus different maximum magnification they can attain}, the ratio between the detection rate of transients in a blue filter to a red filter, $B/R$, should also directly correlate with the abundance of the corresponding stellar populations in the lensed galaxy, at least, to the first order, as the maximum magnification RSGs can attain is lower owing to their larger radius.
\citet{Diego_2024_buffaloflashlights} first investigated how the $B/R$ ratio is related to the relative abundance of RSG and BSG in a lensed galaxy under a simplified scenario, and found that a roughly equivalent abundance of BSG and RSG can reproduce the observed $B/R$ ratio. In light of the strong and different dependencies between BSGs and RSGs with age, the $B/R$ ratio can provide insights into the recent star formation histories (SFHs) of galaxies.

In this paper, we explore how the lensed star $B/R$ ratio depends on star formation parameters and how this ratio can be used in combination with the spectral energy distribution of the galaxy to better constrain its recent SFH. Throughout this manuscript, we adopt the AB magnitude system \citep{Oke_1983}, along 
with standard cosmological parameters: $\Omega_{m} = 0.3$, $\Omega_{\Lambda} = 0.7$, and $H_{0} = 70\ \textrm{km s}^{-1}\, \textrm{Mpc}^{-1}$.

\section{Transient Detection Rate}
\label{sec: method}

\begin{figure}
    \centering
    \includegraphics[width= 0.8\linewidth]{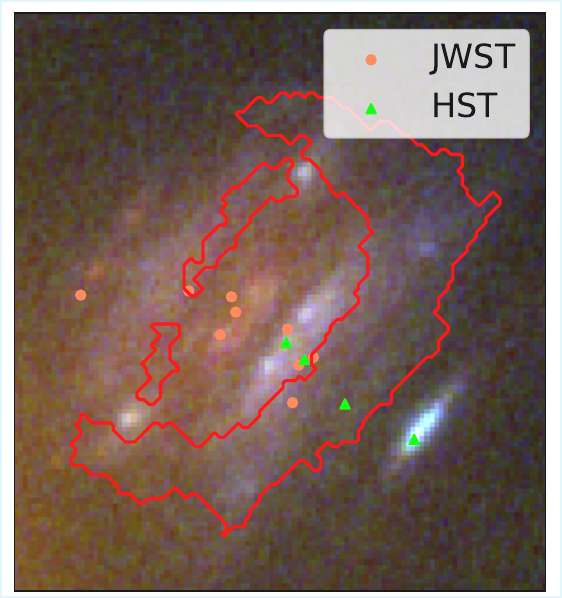}
    \caption{RGB image of the Warhol arc ($z = 0.94$). We show the transients detected in this arc as coral circles \citep[reported by][]{Yan_2023, Williams_2025} and green triangles \citep[reported by][]{Kelly_2022_Flashlights}, respectively. We highlight areas characterized as region 2 by \citet{Palencia_2025_microlensing} in red.}
    \label{fig: warhol}
\end{figure}

As a demonstration, we use a lensed arc in the MACSJ0416 ($z = 0.39$) field known as ``Warhol'' ($z = 0.94$) in which multiple transients have been detected over the last decade \citep{Chen_2019_Warhol, Kelly_2022_Flashlights, Yan_2023}, as shown in Fig.~\ref{fig: warhol}. We adopt the magnification map of Warhol from \citet{Palencia_2025_microlensing}, where for each pixel in which light from Warhol was detected, we have a specific tangential magnification $-10^{4.7}<\mu_{t}<10^{4}$, radial magnification $1.7 < \mu_{r} < 2.3$, and surface mass density of stellar microlenses $30 < \Sigma_{\star} < 115\, M_{\odot}/\textrm{pc}^{2}$.

\begin{figure}
    \centering
    \includegraphics[width=0.95\linewidth]{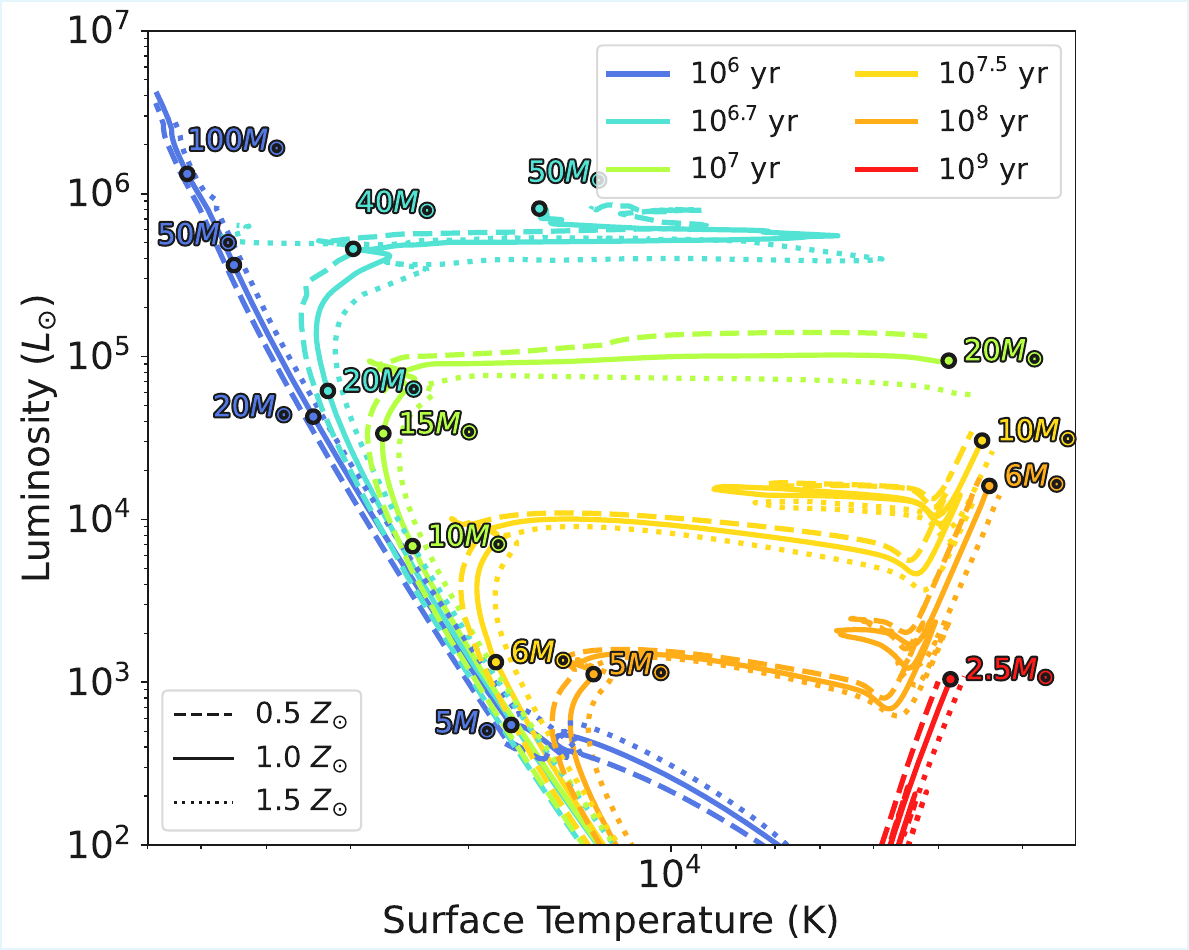}
    \caption{Subsets of MIST age isochrone \citep{MIST} used in our {\tt SPISEA} \citep{SPISEA} stellar population synthesis featuring a range of ages as shown in the legend. The metallicity of each of the subsets is reflected by their linestyle as shown in the bottom legend. All of them have no dust extinction, where the color represents the age of the isochrone as shown in the top legend. For each age isochrone, we also annotate some characteristic masses to indicate the mass-age dependency. }
    \label{fig: CMD}
\end{figure}

We begin by simulating stellar populations having different ages ($t_{age}$), metallicities ($Z_{\star}$), and dust extinctions ($A_{V}$) with using the stellar population synthesis code {\tt SPISEA} \citep[][]{SPISEA} that adopts a MIST isochrone \citep{MIST} and either a \citet{Kroupa_2001} (a power-law slope of 1.3 and 2.3 below and above $1.4\,M_{\odot}$, respectively) or a Top-heavy (power-law slope of 1 for all stellar masses) stellar initial mass function (IMF). Representative of the age isochrones from MIST is shown in Fig.~\ref{fig: CMD} for reference. In our simulations, each stellar population has the same total stellar mass of $10^{6} M_{\odot}$.  

Following the methodology in \citet{Li_2025_IMF} and \citet{Palencia_2025_microlensing}, we compute the expected transient detection rate by calculating the expected number of stars in a stellar population that is brighter than the detection threshold when subject to cluster macrolensing and stellar microlensing. The probability of any stars having some magnification, $p(\mu)$, is characterized by $\mu_t$, $\mu_{r}$, and $\Sigma_{\star}$ \citep{Palencia_2024}. To ensure the unresolved objects in the lensed arcs are individual stars, we demand that they must be detected in the form of transients, such that our sample would not be contaminated by persistent unresolved objects that vary in brightness, such as star clusters \citep[e.g., ][]{Li_2024_Flashlights}. We hence adopt the detectable-through-microlensing (DTM) limit introduced by \citet{Diego_2024_3M} in our calculation, which restricts the individual background lensed stars to be undetectable when subjected to cluster macro magnification only, but becomes temporarily detectable when they are additionally boosted in magnification by microlenses.
Mathematically, the detection rate for any pixel that has $\mu_{t}$, $\mu_{r}$, and $\Sigma_{\star}$, is given by:

\begin{eqnarray}
\label{eqn: detection_rate}
\lefteqn{\mathcal{R}_{f}(\mu_{t}, \mu_{r}, \Sigma_{\star}; t, Z_{\star}, A_{V}) \propto \frac{1}{\mu_{t} \mu_{r}}\int_{-\infty}^{m_{\sigma, f}} \int_{\mu_{\text{min}}= 10^{-1}}^{\mu_{\text{max}} = 10^{4}} \cdot}  \\ \nonumber
& & d\mu \, dm'_{f}\,p(\mu; \mu_{t}, \mu_{r}, \Sigma_{\star}) N_{f}(m-2.5\textrm{log}_{10}\mu; t, Z_{\star}, A_{V})\, ,
\end{eqnarray}

\noindent where $m_{\sigma, f}$ denotes the detection limit in apparent magnitude $m$ at filter $f$. By convolving $p(\mu; \mu_{t}, \mu_{r}, \Sigma_{\star})$ with the stellar luminosity function $N_{f}(m; t, Z_{\star}, A_{V})$ for a stellar population at a given age, metallicity and dust extinction, we can compute the expected number of lensed stars and their brightness distribution after considering all possible lensing magnifications they can attain, $\mu$, through the second integral. The choice of $\mu_{min}$ in this integral is motivated by the $p(\mu)$ resolution \citep{Palencia_2024}, whereas the choice of $\mu_{max}$ is motivated by the maximum size of lensed stars \citep{Li_2025_IMF}. The first integral determines the expected number of stars that are brighter than the detection threshold, given the combined lensing magnification from cluster and stellar microlenses in any random observation. The whole process is computed over every pixel in the image plane representing the arc, requiring us to divide the pixel area by the macro-magnification $\mu_{t}\mu_{r}$ to account for the decrease in source plane area thus the decrease in number of stars. 
We assume a constant surface brightness in the lensed arc for simplicity, where {\it JWST} F200W observation shows that the surface brightness of Warhol only varies by $\sim 40\%$, with diminished variations in regions over which transients are usually detected (near the critical curve).

\begin{figure}
    \centering
    \includegraphics[width=0.95\linewidth]{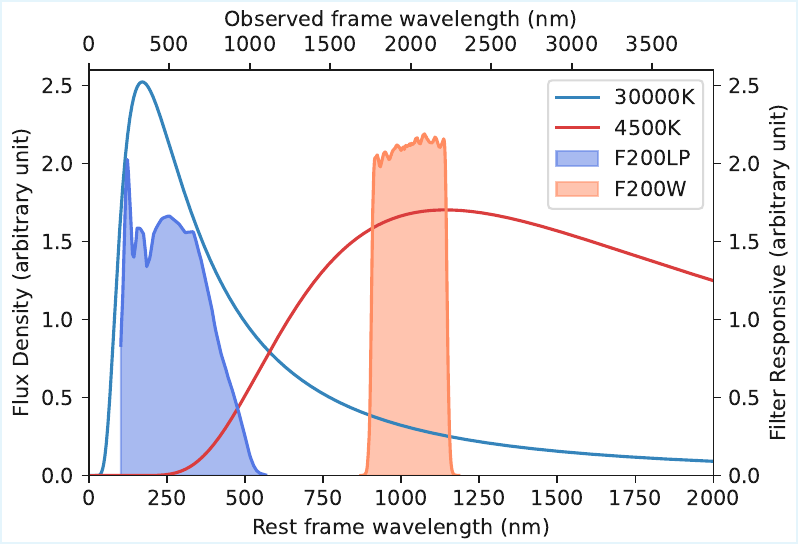}
    \caption{Filter response curve of {\it HST} F200LP (blue shade) and {\it JWST} F200W (coral shade), and the black body spectrum of two stars at $30000\,K$ (blue curve) and $4500\, K$ (red curve) --- characteristic temperature of BSGs and RSGs, respectively --- shown at the rest frame and observed frame ($z = 0.94$) wavelengths. One can see that the two filters best capture the brightest part of the black body of the two classes of stars. }
    \label{fig: BB_FR}
\end{figure}

\begin{figure}
\centering
    \includegraphics[width=0.95\linewidth]{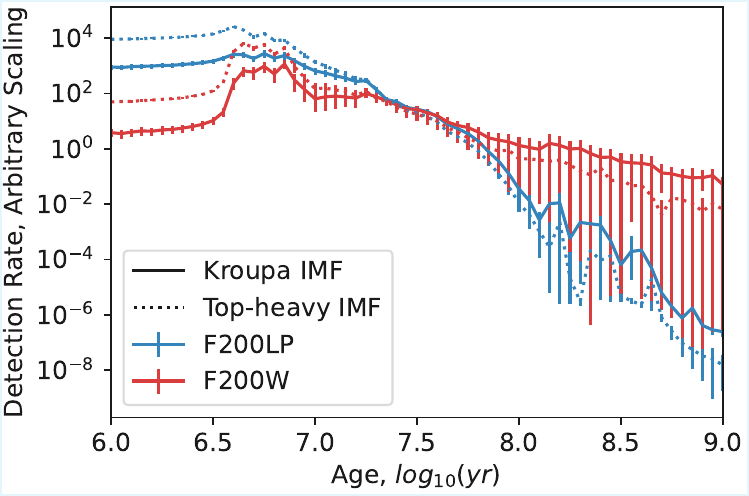}
    \caption{Transient detection rate per unit mass for stellar populations at different ages with metallicity of $0.5 Z_{\odot}$ and $A_{V} = 0$, calculated with Eq.~\ref{eqn: detection_rate} under the magnification map of Warhol from \citet{Palencia_2025_microlensing}. Blue curves show the detection rate in the {\it HST} F200LP filter (detection limit of $30\,m_{AB}$); red curves show the detection rate in the {\it JWST} F200W filter (detection limit of $29.7\,m_{AB}$). Solid and dotted lines show the case for a Kroupa IMF and a top-heavy IMF, respectively. }
    \label{fig: detection_rate}
\end{figure}

We select two characteristic filters that best capture the light from BSGs and RSGs at $z \approx 1$. For the blue filter, we choose the {\it HST} WFC3/UVIS F200LP as it spans the rest-frame UV through optical ($100\,$nm to $540\,$nm) and has the highest sensitivity to BSGs; for the red filter, we choose the high-throughput {\it JWST} NIRCAM F200W as it covers rest-frame NIR ($850\,$nm to $1.2\,\mu$m) where RSGs are the brightest. Owing to the high sensitivity of this filter, it is used in many deep programs. The filter response curves and characteristic blackbody spectra of BSGs and RSGs are shown in Fig.~\ref{fig: BB_FR}. We adopt a detection limit of $30\,m_{AB}$ for F200LP, as motivated by the {\it HST} Flashlights survey \citep{Kelly_2022_Flashlights} and $29.7\, m_{AB}$ for F200W, as motivated by the {\it JWST} PEARLS program \citep{Windhorst_2023_PEARLS, Williams_2025}. Notice that BSGs can sometimes be detected in F200W, and conversely, RSGs in F200LP. The likelihood of such detections, however, is low given the bandpasses of the respective filters compared with the blackbody spectra of the different stars. An example of the transient detection rate in these two filters is shown in Fig.~\ref{fig: detection_rate}, for which those using other combinations of metallicity and dust are shown in Fig.~\ref{fig: detection_rate_all}.

The transient $B/R$ ratio for any given stellar population is then just the ratio between the detection rate summed over all the $n$ lens model pixels that represent the arc in the two filters:

\begin{equation}
    B/R  = \sum^{n}_{i}\mathcal{R}_{F200LP}(\mu_{t, i}, \mu_{r, i}, \Sigma_{\star, i})/\sum^{n}_{i}\mathcal{R}_{F200W}(\mu_{t, i}, \mu_{r, i}, \Sigma_{\star, i})
\end{equation}

\begin{figure*}
\centering
    \includegraphics[width=\linewidth]{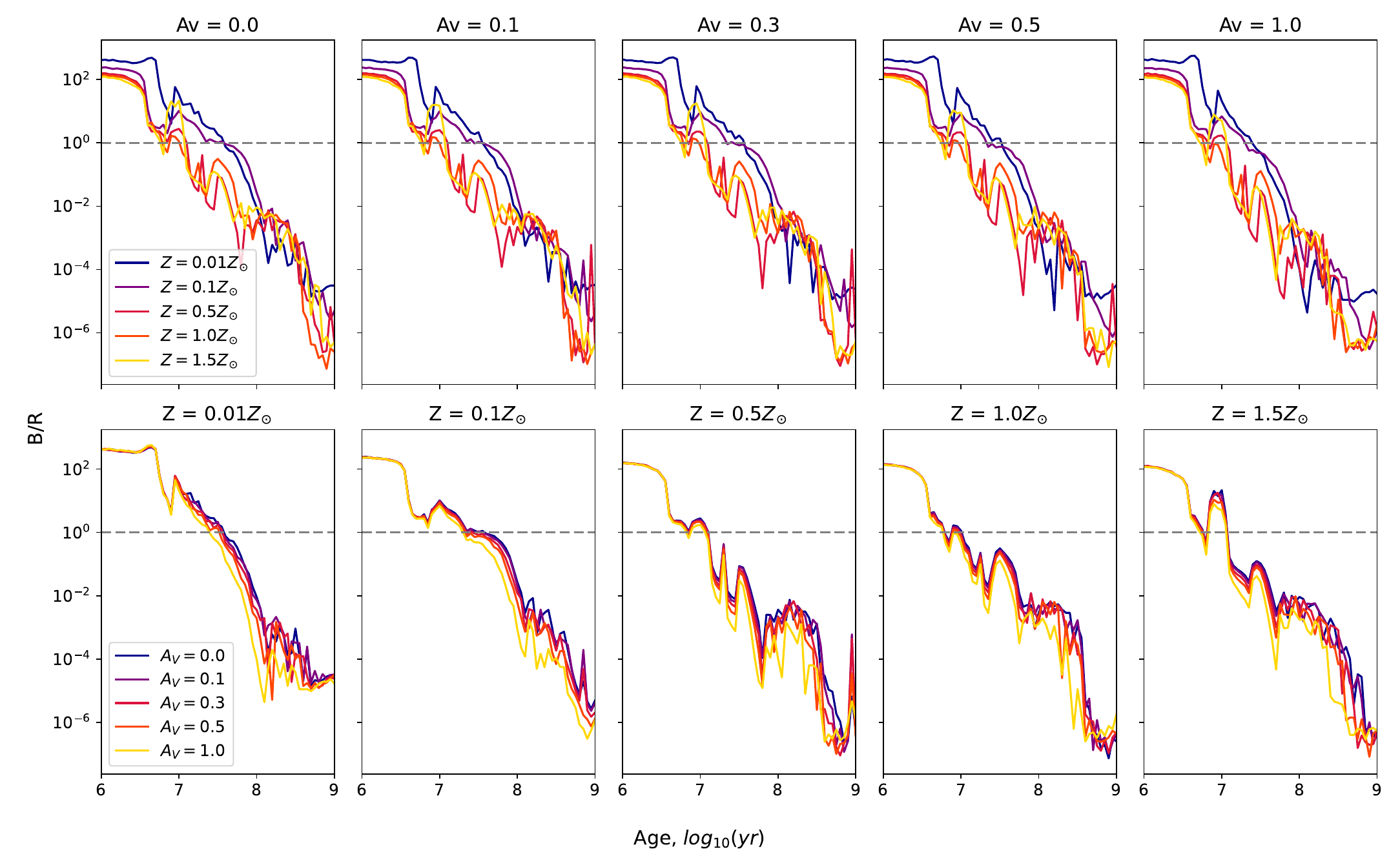}
    \caption{Expected $B/R$ ratio calculated based on the F200LP and F200W transient detection rate in Warhol arc shown in Fig.~\ref{fig: detection_rate_all}, with a Kroupa IMF. Each of the panels shows the case for one of the dust extinctions (upper row) or one of the metallicities (lower row) as indicated in the subtitles. The color of each curve indicates the metallicities (upper row) or dust extinction (lower row). The gray dashed line denotes $B/R = 1$ for reference. The fluctuation in ages earlier than $\sim 50\,$Myr are significant and arise from intrinsic variation due to stellar evolution, whereas those later than $\sim 50\,$Myr are arising from sampling uncertainty given the low transient detection rate at these ages. We do not show the error bars for visualization purposes. }
    \label{fig: BR_Ratio}
\end{figure*}

\noindent We show dendence of the $B/R$ ratio with age for stellar populations having different metallicities, $Z_{\star}$, and for each $Z_{\odot}$ subject to different dust extinctions, $A_{V}$, in Fig.~\ref{fig: BR_Ratio}.

\section{Properties of the $B/R$ ratio}
\label{sec: properties}

As shown in Fig.3, the transient detection rate per unit mass at $z \approx 1$ in both the blue ({\it HST} F200LP) and red ({\it JWST} F200W) filters is highest for star formation episodes over the past $\sim 10\,$Myr (notice that the peak in transient detection rate is slightly delayed in the Red filter, as massive stars take time to evolve into RSGs). Beyond $\sim 10\,$Myr, the transient detection rate decreases quickly with increasing age as most BSGs and a significant fraction of RSGs ($\gtrsim 10^{4}L_{\odot}$) only survive up to $\sim 10\,$Myr as shown in Fig.~\ref{fig: CMD}. After this age, better still, the transient detection rate is dominated by less massive evolved stars. These stars, however, contribute much less to the transient detection rate than BSGs/RSGs, given their lower intrinsic brightnesses as shown in Fig.~\ref{fig: CMD}, requiring a higher (less likely) magnification to be detected as a transient. 

For a coeval stellar population having age $>10\,$Myr, the transient detection rate decreases more slowly with age in the R than B filters. This behaviour translates to a decreasing $B/R$ ratio with age as shown in Fig.~\ref{fig: BR_Ratio}. As an example, for a stellar population at the youngest age of $\lesssim 3\,$Myr, the detection rate is significantly higher in the blue filter, thus yielding a high B/R ratio of $\sim 10^{2-3}$. At such young ages, the DTM stars are mostly BSGs, as demonstrated by the blue isochrone in Fig.~\ref{fig: CMD}, while less massive stars need more time to evolve into the RSG phase. Although BSGs are so bright that they can sometimes be detected also in the red F200W filter, they primarily contribute to the detection rate in the blue filter, as shown in Fig.~\ref{fig: detection_rate}. On the other hand, red DTM stars (most likely, RSGs) only become appreciable at a slightly older age of $\gtrsim 5\,$Myr (the cyan isochrone in Fig.~\ref{fig: CMD}) as corresponding to the time taken by intermediate-mass stars to evolve away from the main sequence. The $B/R$ ratio thus drops quickly once massive stars start to enter the RSG phase. 
At even older ages of $\gtrsim 5-40\,$Myr (depending on the metallicity/dust extinction), the most massive and thus brightest BSGs are gone, with few of the dimmer BSGs surviving as shown by the green isochrone in Fig.~\ref{fig: CMD}. During this time, the $B/R$ ratio falls below unity as reflected in Fig.~\ref{fig: BR_Ratio} as the only remaining DTM stars are relatively dim RSGs alongside other red evolved stars. This trend continues at even older ages, during which the $B/R$ ratio fluctuates owing to effects related to metallicity and dust extinction as explained next.

As can be seen in Fig.~\ref{fig: BR_Ratio}, the dependence of the $B/R$ ratio on metallicity is weaker than on age, but still could promote or demote the $B/R$ ratio up to factors of $\sim 10 - 100$ depending on the age of the stellar population. There are a couple of factors that determine how metallicity affects the $B/R$ ratio. The first is the metal absorption in stellar atmospheres, for which there are more metal absorption lines in the B than R filters, such that an increase of metallicity would generally decrease the $B/R$ ratio, as could be seen in Fig.~\ref{fig: BR_Ratio}. Furthermore, as shown by the isochrones in Fig.~\ref{fig: CMD}, stars with high metallicities (indicated by dotted lines) are cooler (thus redder) compared with those with lower metallicity (indicated by dashed lines). Another complicated effect would be the role of metallicity in stellar evolution and, therefore, the time stars of different masses spend at different stages, which is more age-specific. For example, at super-solar metallicity (right-most panel, Fig.~\ref{fig: BR_Ratio}), the increase in stellar opacity owing to metal absorption inflates the star and leads to increased mass loss via stellar wind \citep{Vink_2001}. With the outer envelope stripped away, the massive stars cannot evolve into RSGs and remain as BSGs, leading to a dip in the transient detection rate in the R filter at $\sim 10\,$Myr, therefore explaining the bump in the $B/R$ ratio at this age. These effects depend on the isochrone adopted \citep{MIST}, and we do not analyze all the trends in further detail in this work.

Compared with metallicity, dust extinction has an even more minor effect on the $B/R$ ratio. Increasing the dust extinction results in more blue light being absorbed, making the observed stellar population redder in general. This increase leads to the detection of more red transients and suppresses the $B/R$ ratio at any age. Such an effect is more appreciable after an age of $\sim 10\,$Myr --- the DTM stars at such age are much dimmer, and the dust extinction would make them even dimmer in the B filter, thus further lowering the $B/R$ ratio. 

Last but not least, we found that the $B/R$ ratio varies relatively weakly with the two adopted IMFs. We do not show the corresponding results for a top-heavy IMF in Fig.~\ref{fig: BR_Ratio}, as the changes induced compared with a Kroupa IMF (as shown in Fig.~\ref{fig: BR_Ratio}) are much weaker than the observed variation with age. A shallower IMF gives rise to more blue transients at ages less than $3\,$Myr, with a fractional change in the $B/R$ ratio of $\lesssim 20\%$. This difference in the $B/R$ ratio for the different stellar IMFs diminishes ($\lesssim 3\%$) at older ages, as Kroupa has a similar slope as the Top-heavy IMF for low- to intermediate-mass stars.

\section{Insights into star formation}

In Sec.~\ref{sec: properties}, we showed how the B/R ratio changes dramatically with the age of a stellar population, to a lesser extent, metallicity, and finding that this ratio is quite insensitive to the assumed IMF (whether Kroupa or Top-heavy) as well as dust extinction. As a consequence, the B/R ratio can be used as a powerful probe of the recent star-formation history of galaxies that are multiply-lensed by galaxy clusters. In this section, we show that the best-fit synthetic stellar population to the spectral energy distribution (SED) of the Warhol arc predicts incompatible B/R ratios for this galaxy, and vice versa, no matter whether we adopt a single burst or a more extended duration of star formation. We then show how fitting to both the B/R ratio and SED simultaneously improves our understanding of the recent star-formation history of this galaxy, predicting a quite different stellar mass formed over the past $\sim50\,$Myr compared to that implied when fitting either constraint alone.

\begin{figure}
    \centering
\includegraphics[width=0.9\linewidth]{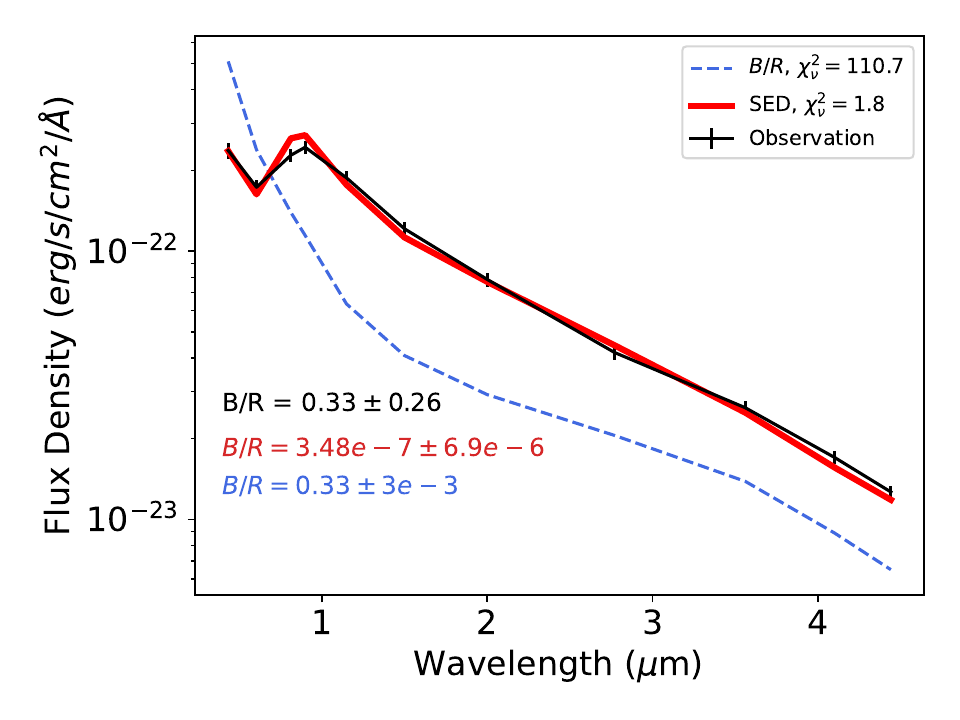}
    \caption{SED of Warhol arc \citep[black curve, from][]{Palencia_2025_microlensing}. The best fit (red curve) corresponds to a single star burst model at an age of $\sim 350\,$Myr with $Z_{\star} = 0.32\pm0.12\,Z_{\odot}$ and $A_{V} = 0.91\pm0.04$. This model predicts $B/R = 10^{-7}$, orders of magnitude from the observed $B/R = 0.33\pm0.26$. Among all the models that can reproduce the observed $B/R$ ratio well, we show the best fit SED as a blue curve. This model corresponds to a single star burst at an age of $\sim 80\,$Myr with $Z_{\star} = 0.8\pm0.4\,Z_{\odot}$, and $A_{V} = 0.5\pm0.3$. We adopted a Kroupa IMF in both models. }
    \label{fig: SED_fit}
\end{figure}

\subsection{Single starburst}

As proof-of-concept, we begin the simplest possible case where there is only one coeval population of young stars. The observed $B/R$ ratio of the Warhol arc is $0.33 \pm 0.26$ (\cite{Kelly_2022_Flashlights, Yan_2023}, summarized by \cite{Williams_2025}). In the case of a single stellar population, we deduce from Fig.~\ref{fig: BR_Ratio} that this arc must have experienced a starburst $\sim 5-40\,$Myr ago, depending on the assumed metallicity. As we show in Fig.~\ref{fig: SED_fit}, however, the observed SED \citep[black curve, from][]{Palencia_2025_microlensing} is not consistent with a starburst over the aforementioned age range, for which the blue curve shows the predicted SED for an age of $12\,$Myr (that which minimizes differences with the observed SED, but for which the reduced $\chi^{2} \approx 100$). Conversely, fitting a single starburst generated by {\tt SPISEA} to the observed SED of Warhol, we find a best-fit age of $\sim 350\,$Myr (requiring $0.3 Z_{\odot}$ and $A_{V} = 0.9$) for which the reduced $\chi^{2}$ is 1.8 (red curve in Fig.~\ref{fig: SED_fit}). The predicted $B/R$ ratio at this age, however, is of order $10^{-7}$, in dramatic contradiction with the observed $B/R$ ratio.

The corresponding Markov-Chain Monte-Carlo (MCMC, see details in Section~\ref{sec: mcmc}) sampling of the constraints imposed by either the $B/R$ ratio or SED alone on the properties of a single starburst is shown in Fig.~\ref{fig: mcmc}. As can be seen, regardless of whether we adopt a Kroupa or Top-Heavy IMF, there is a $\sim 4\sigma$ tension between the age of the starburst inferred from the $B/R$ ratio (blue contours) or SED (red contours) alone. This contradiction clearly shows that the Warhol arc must have a more complicated SFH so as to reproduce the SED and $B/R$ ratio simultaneously.

\subsection{Continuous star formation}
\label{sec: non-param}

We now consider a more complicated star-formation history, in which the star-formation rate is allowed to be freely fit in five different age bins spanning $1 - 5\,$Myr, $5 - 10 \,$Myr, $10 - 50\,$Myr, $50 - 250\,$Myr, and $250\,$Myr $-$ $1\,$Gyr. We adopt a Kroupa IMF, having shown earlier that a Top-heavy IMF makes little change to the predicted B/R ratio.

\begin{figure}
    \centering
\includegraphics[width=0.9\linewidth]{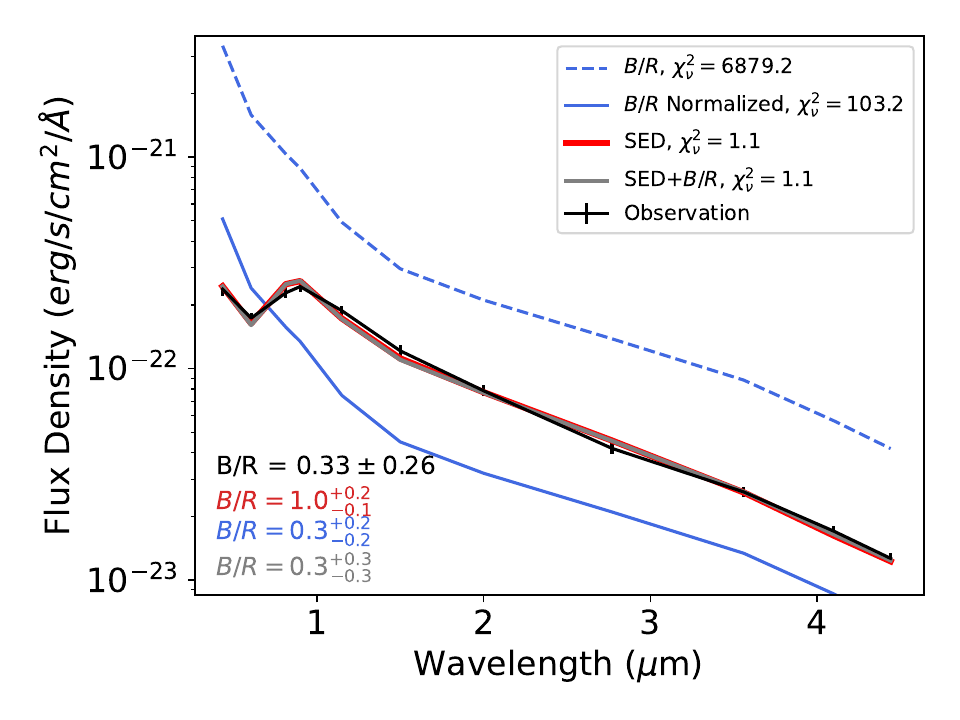}
    \caption{Same as Figure~\ref{fig: SED_fit}, but we use a non-parametric SFH model instead of a single star burst to fit the data. Again, the SFH that is constrained by the SED (red curve) does not predict the $B/R$ ratio well (predicted $1.0^{+0.2}_{-0.1}$, versus observed $0.33\pm0.26$), while that is constrained by the $B/R$ poorly predicts the SED (blue curves). Notice that the $B/R$ ratio is not sensitive to the underlying stellar mass; thus, the SED inferred from $B/R$ deviates significantly from the observed SED (dotted blue curve). We enforce a normalized version of the same SED (solid blue curve), where the relative star formation rate in each age bin of the underlying SFH is invariant, to acquire the best fit to the observed SED. Nevertheless, imposing both the SED and $B/R$ as constraints infers a solution that fits both of the constraints to a sensible degree, which signifies the importance of including the $B/R$ ratio as extra constraints to better capture the true SFH. The best-fit solution has a metallicity of $0.44\pm0.21 Z_{\odot}$ and a dust extinction of $0.91\pm0.07$, respectively. }
    \label{fig: SED_fit_nonparam}
\end{figure}

We show the best-fit SED in Fig.~\ref{fig: SED_fit_nonparam}, demonstrating a $2\sigma$ tension between the resulting predicted and observed B/R ratios ($1.0^{+0.2}_{-0.1}$ versus $0.33\pm0.26$) when the SED is used as the sole constraint on the SFH.  Similarly, we also demonstrate the resultant tension between the predicted (blue curve) and observed (black curve) SED for the best-fit SFH model when constrained by the B/R ratio alone. As the B/R ratio does not depend on the stellar mass formed, the model SED plotted in blue is that which best reproduces the measured SED given the best-fit SFH to the B/R ratio.  Evidently, increasing the complexity of the SFH does not resolve the tension between best-fit model SFHs as constrained by the SED or B/R ratio individually.

\begin{figure}
    \centering
\includegraphics[width=0.9\linewidth]{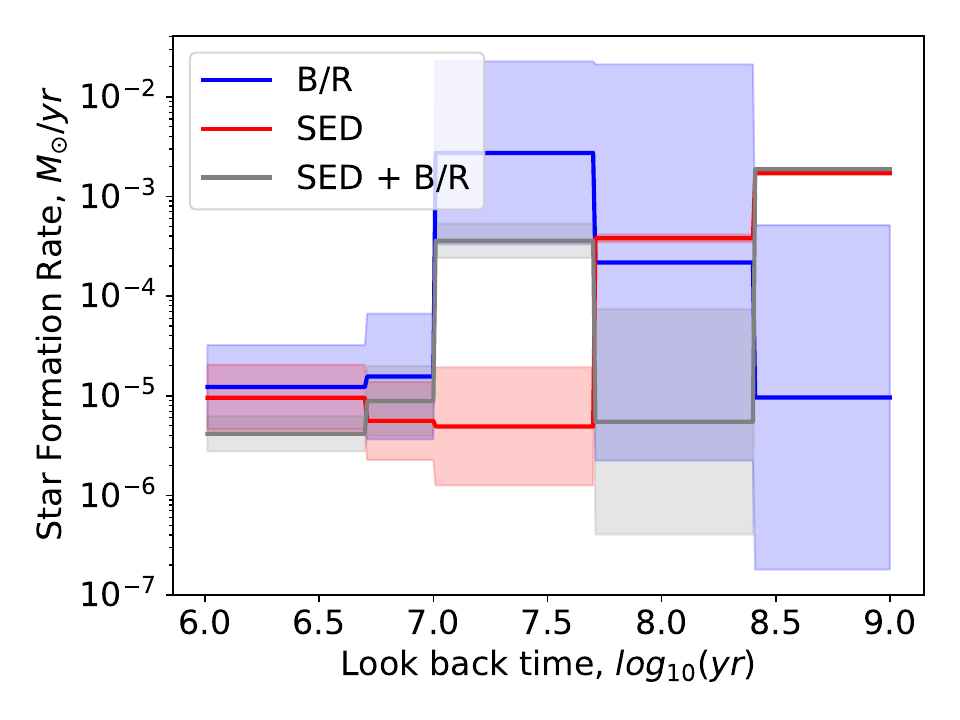}
    \caption{Corresponding best-fit SFH models of the Warhol arc that generate the SED and $B/R$ ratio as shown in Fig.~\ref{fig: SED_fit_nonparam} with the band representing the $1\sigma$ tolerance from MCMC analysis. One can see that the inclusion of the $B/R$ ratio as constraint on top of the SED give rise to a different SFH with more star formation over the past $10-50\,$Myr (factor of $\sim 60$ with $\gtrsim 3\sigma$ significance), and fewer star formation over the last $50-250\,$Myr (factor of $\sim 60$ with $2\sigma$). }
    \label{fig: SFH_nonparam}
\end{figure}

In light of the persistent tension, we conduct another MCMC analysis to test whether there is any model SFH, parameterised in the manner described above, that can provide an acceptable fit to both the $B/R$ ratio and SED simultaneously. We are able to find such a solution for which the model SFH is shown in Fig.~\ref{fig: SFH_nonparam}, to be compared with the best-fit model SFHs when fitting for either the SED or B/R ratio alone. The corresponding predicted SED is shown by the gray curve in Fig.~\ref{fig: SED_fit_nonparam} with the MCMC result shown in Fig.~\ref{fig: mcmc_nonparam}. The metallicity inferred in this solution is $0.44 \pm 0.21 Z_{\odot}$ with dust extinction of $0.91\pm0.07$. As can be seen, the model SED closely resembles the observed SED (black curve). Furthermore, despite the different implied SFHs as shown in Fig.~\ref{fig: SED_fit_nonparam}, the model-predicted SED constrained by the SED and $B/R$ ratio simultaneously is very similar to that constrained solely by the SED. At the same time, when constrained by both the SED and $B/R$ ratio simultaneously, the $B/R$ ratio is now also reproduced to well within $1\sigma$. When constrained by both the SED and $B/R$ ratio simultaneously, the stellar mass formed over the past $\sim 50\,$Myr is $\sim 12300^{+7200}_{-4500} M_{\odot}$ ($0.9_{-0.3}^{+0.5}\%$ of the total stellar mass), two orders of magnitude higher compared with that implied by the corresponding SFH models constrained by the SED alone ($\sim 260^{+580}_{-150} M_{\odot}$, $0.02^{+0.04}_{-0.01}\%$ of the total stellar mass) with a $\sim 3\sigma$ tension.

We also compare the Bayesian Information Criterion (BIC) to see which of the two aforementioned models is preferred. The BIC is given as:

\begin{equation}
    BIC = k\, \textrm{ln}(N) - 2\, \textrm{ln}(L_{max}) ,
\end{equation}

\noindent where $k$ is the number of model parameters, $N$ the number of data points, and $L_{max}$ the maximum likelihood of the given model. The best-fitting model that is constrained only by SED has a BIC of 45, and the best-fitting model that is constrained by both SED and $B/R$ ratio has a BIC of 30. This means that there is very strong evidence (difference in BIC $> 10$) that the latter model is better at reproducing the observations.

The main difference between the SFH constrained by SED alone and by SED and $B/R$ ratio simultaneously is that the latter one has a significantly higher SFR (factor of $\sim 60$, $\gtrsim 2\sigma$) in the third age bin of $10-50\,$Myr, and a lower SFR at the fourth age bin of $50-250\,$Myr ($2\sigma$). Since a certain amount of UV bright stars (only formed in the last $\sim 10\,$Myr) is required to fit the rest-frame UV part of the SED, the only way to reduce the over-predicting $B/R$ ratio is to form more red stars that boost the detection rate in the R filter. The age bin $10-50\,$Myr contains a relatively high abundance of RSGs compared to BSGs, whereas an increased SFR in this bin suppresses the $B/R$ ratio (c.f. Fig.~\ref{fig: BR_Ratio}, middle panel) to fit the observation without sacrificing the quality of fits towards the SED. On the other hand, the SFR is lower in the fourth bin, possibly a result of compensating for the increased total fluxes due to the increased SFR in the third age bin. The transient detection rate between ages of $50-250\,$Myr is expected to be very low (c.f. Fig.~\ref{fig: detection_rate}), such that the variation of SFR in this bin would not affect the inferred $B/R$ ratio at all. This result also hints towards that perhaps the recent star formation in the Warhol galaxy is rather bursty in nature, and the $B/R$ ratio could be constraining the burstiness.

\subsection{Spatial variation in star formation history}
\label{sec: region2}

One of the benefits of using the $B/R$ as a constraint of the SFH is that it has extremely high spatial sensitivity down to the pixel level. 
The Warhol arc has varying colors over different regions -- which is a strong indicator of varying SFHs in recent times over different parts of the galaxy. Among the six regions of Warhol as characterized by their photometric color \citep{Palencia_2025_microlensing}, only region 2 (as highlighted in Fig.~\ref{fig: warhol}) has transient detection in both B and R filters with a $B/R$ ratio of $0.3\pm0.4$. Since we only have a measurement of $B/R$ at region 2, we cannot investigate how the inclusion of $B/R$ ratio as a constraint imposes different SFH compared with those constrained only by SED in different parts of the arc -- instead, we can compare the local SFH confined to region 2 with the globally inferred SFH.

\begin{figure}
    \centering
\includegraphics[width=0.9\linewidth]{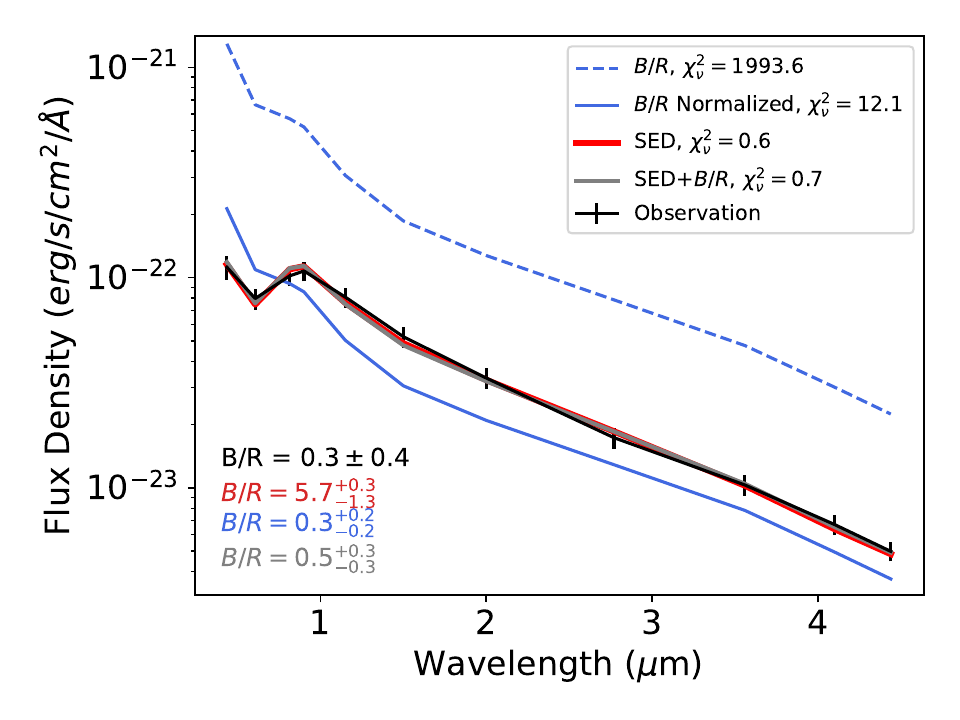}
    \caption{Same analysis in Fig.~\ref{fig: SED_fit_nonparam}, but on region 2 of Warhol (highlighted in Fig.~\ref{fig: warhol}), as characterized by \citet{Palencia_2025_microlensing}. The best fit model that is constrained by both SED and $B/R$ ratio has a metallicity of $0.12\,Z_{\odot}$, and a dust extinction of $A_{V} = 0.53 $, agreeing with the SED fitting result in \citet{Palencia_2025_microlensing} to within $1\sigma$. }
    \label{fig: Region2_SED}
\end{figure}

\begin{figure}
    \centering
\includegraphics[width=0.9\linewidth]{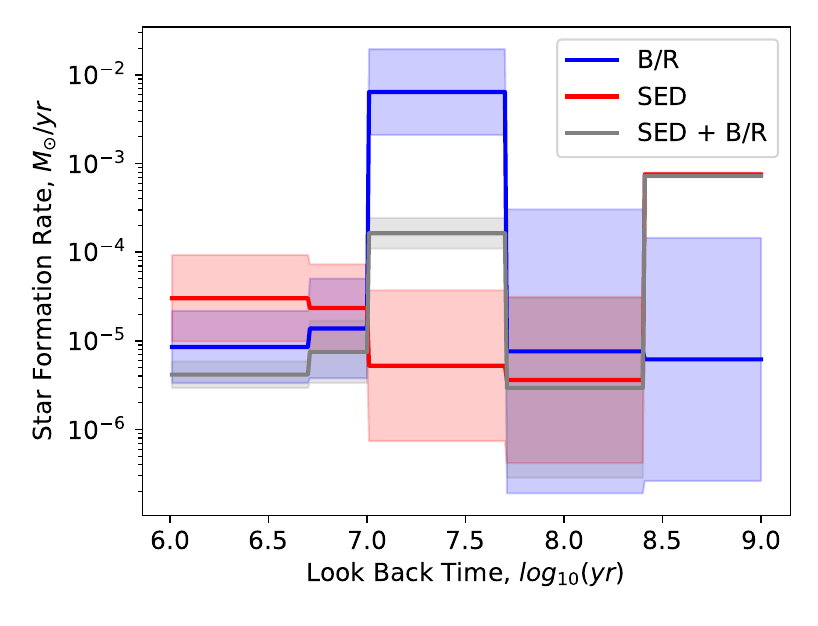}
    \caption{SFH model corresponding to the analysis in Fig.~\ref{fig: Region2_SED}, but on region 2 of Warhol, as characterized by \citet{Palencia_2025_microlensing}. Again, the inclusion of $B/R$ ratio as a model constraint on top of the SED predicts a different SFH. The disagreement between the two solutions of SFH when the $B/R$ ratio is incorporated as a constraint, however, is more drastic in the case confined to region 2 than in the case for the whole Warhol arc. This demonstrates the spatial variation of SFH in the lensed galaxy, and the constraining power of $B/R$ at different regions. }
    \label{fig: Region2_SFH}
\end{figure}

We apply the same calculation described earlier in Sec~\ref{sec: non-param} on region 2 and show the results in Fig.~\ref{fig: Region2_SED} and Fig.~\ref{fig: Region2_SFH}, where the MCMC result can be found in Fig.~\ref{fig: mcmc_nonparam_R2}. When SED is the only constraint, the $B/R$ ratio is poorly predicted ($5.7^{+0.3}_{-1.3}$ versus $0.3\pm0.4$) with worse tension ($>3\sigma$) than the case for the entire Warhol arc. Again, when $B/R$ is incorporated as an additional constraint, the tension is resolved. One can tell from Fig.~\ref{fig: Region2_SFH} that the SFH inferred is very different when the SED alone (red), or the SED and $B/R$ ratio are used as constraints together (gray). The recent SFR ($1-10\,$Myr) is suppressed by a factor of $\sim 10$ with $\sim 1-2\sigma$, and the SFR at $10-50\,$Myr ago is boosted by a factor of $\sim 20$ with $> 3\sigma$ upon the inclusion of $B/R$ ratio as constraints. The total stellar mass formed thus increases from $450^{+1300}_{-200} M_{\odot}$ ($\sim 0.1^{+0.3}_{-0.1}\%$ of the total stellar mass) to $6600^{+3300}_{-2200} M_{\odot}$ ($\sim 1.4^{+0.7}_{-0.4}\%$ of the total stellar mass).
We again compare the BIC of the two models, finding that the former model has a BIC of 26 and the latter has a BIC of 23. The difference between BIC decreases, however, the model that is constrained by both SED and $B/R$ is still moderately preferred over the one that is only constrained by the SED.

Notice that region 2 is a subset of Warhol -- the fact that the inferred SFH of region 2 forms more stellar mass in the last $\sim 50\,$Myr than that inferred for the whole Warhol arc indicates that there are severe degeneracies in the SED fitting \citep[e.g., long-known stellar mass and metallicity degeneracy, ][]{Worthey_1994}, especially when the difference among the different regions of galaxies are smoothed out during the computation of SFH of the galaxy as a whole as previously found in literature \citep[e.g., ][]{Sorba_2018, Gimenez-Arteaga_2024}. This stresses the importance of having spatially varying information and highlights the possible role of $B/R$ in improving the constraints on SFH upon lensed galaxies that are spatially resolved.

Also, our inferred SFH of region 2 is different from that inferred in \citet{Palencia_2025_microlensing}, even when the exact same SED is used as the constraint. The main driver behind this discrepancy is the model complexity, as \citet{Palencia_2025_microlensing} has only explored an exponential decay model that does not freely allow for recent star formation. Although all the SFHs have the highest star formation rate at $\sim 250\,$Myr - $1\,$Gyr, our solution prefers some elevated star formation rate during the last $\sim 50\,$Myr when a non-parametric SFH model is constrained by the SED, where the exact time of the elevated star formation can be better constrained when the $B/R$ ratio is used as a constraint. As \citet{Palencia_2025_microlensing} did not compute the detection rate in {\it HST} F200LP, we cannot further comment on the difference between the different underlying SFH models adopted and how they predict the $B/R$ ratio when only the SED is used as the constraint.

\section{Discussion}

\subsection{Merits of $B/R$ as constraints on SFH }

Further constraints on the SFH can be obtained from spectroscopy targeting emission lines. \citet{Williams_2025} conducted SED fitting on Warhol, which also included the emission line fluxes (H$\alpha$, H$\gamma$, H$\delta$, [O III], [O II], and [Ne III]) of Warhol as constraints. Similarly to us, they use a non-parametric SFH with almost the same set of age bins (except that they have two bins of $1-2\,$Myr and $2-5\,$Myr, instead of a single bin of $1-5\,$Myr as we do; and they have an extra bin for $\gtrsim1\,$Gyr). Based on the inferred SFH, they predict a $B/R$ ratio of $0.07^{+0.07}_{-0.04}$ when adopting an IMF slope of $2.3$ (same as Kroupa IMF at $M_{\odot} \geq 1.4$). This is within $\sim 1\sigma$ of the observed value of $0.33\pm0.26$. 
Since the SFH shown in \citet{Williams_2025} is not corrected for lensing magnification, we can only compare the relative SFR at different age bins between our SFHs by normalizing the SFR to the first star formation bin. We find their inferred relative SFR to be consistent with ours in each age bin to within $ 2\sigma $ except the last age bin of $\sim 250\,$Myr$-1\,$Gyr, which is not associated with either emission line strength, or transient detection (thus $B/R$ ratio).

Although in this case, the most recent relative SFR inferred by either emission line strength or the $B/R$ ratio on top of SED is similar, it is nevertheless worth noticing that the SFH constrained by emission lines, such as the Balmer lines, could be biased towards the most recent SFR within the last $\sim 10\,$Myr \citep{Velazquez_2021}, which is contributed primarily by star-forming regions that are strongly emitting in UV. On the other hand, the $B/R$ ratio can trace star formation up to $\sim 50\,$Myr, until the detection rate per stellar mass begins to decrease significantly (c.f. Fig.~\ref{fig: detection_rate}). 

One of the remaining tests for the merits of the $B/R$ ratio as a constraint of the SFH is whether our solution can reproduce the measured emission line fluxes, or, ultimately, to what extent we can constrain the most recent star formation history when all SED, emission line strength, and $B/R$ ratio are used as constraints together.
The different sensitivity of these two constraints could allow one to better constrain the recent SFR, especially regarding the burstiness of star formation. Limited to our customized code that constrains SFH with $B/R$ ratio alongside the SED, we cannot predict the emission line strength and see how well our best-fitting solution reproduces the emission line strength reduced in \citet{Williams_2025} under the current framework. A full exploration is deferred for future work.

\subsection{Differential dust extinction}

One could argue that differential dust extinction in galaxies \citep[e.g., ][]{Mawatari_2025} could also result in a change in $B/R$ across regions as shown in Fig.~\ref{fig: BR_Ratio}, without altering the SFH. $A_{V}$ is always a free parameter in our MCMC analysis with a flat prior, whereas the two solutions (constrained by SED only, and both SED and $B/R$ ratio) have a slightly different $A_{V}$ of $0.85 \pm 0.07$ and $0.73\pm 0.12$, respectively. Although the degree of freedom is well given, the fact that both solutions still prefer a similar amount of dust extinction means that such a dusty solution is required to explain the SED. Thus, the solution prefers to vary the SFH in order to resolve the $B/R$ ratio without sacrificing the quality of fits towards the SED.

Also, despite the fact that one cannot rule out extremely high dust extinction, the current solution of $A_{V} \approx 0.8$ is already very dusty. And, even with such a level of dust extinction, the solution that is only constrained by SED also over-predicts the $B/R$ ratio. It is not impossible that the transient stars detected are more heavily dust-obscured \citep[for example, from their own stellar wind, e.g., ][]{Tambovtseva_2008} compared with the other stellar population and therefore way redder than what is expected from our stellar population synthesis. However, such a scenario can only be confirmed upon spectroscopic analysis of these lensed stars in future detections; hence, the variation in SFH might be the simplest, viable solution under the principle of Occam's razor.

\subsection{Uncertainties}

Throughout the whole discussion, we have been assuming that the magnification map we have adopted is a truthful representation of the real lensing magnification. The two primary factors that could have affected the lensing magnification are the macro lens model itself \citep[e.g., ][]{Perera_2025}, as well as perturbations in the magnification distribution arising from, for example, substructures in dark matter haloes \citep[e.g., ][]{Moore_1999, Dai_2018}.

The Warhol arc contains many unresolved features, presumably, star clusters, that are multiply-lensed. These features form anchor points to allow one to pinpoint where the macro model's critical curve should cut through \citep{Broadhurst_2025} with little room for uncertainty. The lens model adopted by \citet{Palencia_2025_microlensing} is the \citet{Diego_2024_MACSJ0416_lensmodel} lens model, where one can see that the critical curve closely resembles the expectation demonstrated in the last figure in \citet{Broadhurst_2025}. In such a case, the region where the macro-magnification can be wrongly estimated is rather small. Even though the critical curve is shifted by a bit, the high magnification region ($\mu_{m} \gtrsim 10^{2}$) that contributes the most to the detection rate \citep[see][]{Palencia_2025_microlensing, Li_2025_IMF} remains inside the Warhol arc as predicted by the \citet{Diego_2024_MACSJ0416_lensmodel} lens model. Far regions at the two sides of the critical curve of the Warhol arc are sufficiently far away from the critical curve and have low magnification ($\sim 1-10$) \footnote{$\mu_{m}$ is inversely proportional to the distance to the critical curve}, such that the predicted magnification therein would not change a lot even if there is a small shift in the position of the critical curve. These low macro-magnification regions barely contribute to the detection rate \citep{Li_2025_IMF, Palencia_2025_microlensing} and hence, the $B/R$ inferred should not change a lot with the choice of lens model.

For substructures, \citet{Palencia_2025_microlensing} studied the role of non-luminous cold dark matter subhaloes (or millilenses) with masses of $\sim 10^{6-8} M_{\odot}$ in the detection rate of transients in the Warhol Arc. The main finding is that although the existence of substructures redistributes the magnification, they primarily affect the spatial distribution of transient events \citep[see also][]{Williams_2024, Broadhurst_2025}, and only affect the detection rate of transients down to a $\sim 1\%$ level. Propagating this subtle uncertainty barely affects the resulting $B/R$ ratio, which is currently dominated by Poisson noise. In such a case, our inference of SFH based on the $B/R$ ratio should be invariant under the effect of substructures. Nevertheless, there are more possible cases of substructures to be considered \citep[e.g., density modulations from wave-like dark matter, ][]{Amruth_2023}, and a full exploration is required in future works.

\section{Conclusion}

In this paper, we have found that the ratio of transient detection in two characteristic filters, {\it HST} F200LP, and {\it JWST} F200W, which best capture the light of blue and red supergiants at $z \approx 1$, is a very sensitive probe of the age of the underlying stellar population in lensed galaxies. Based on this fact, we demonstrated that $B/R$ ratio, the transient detection rate in the two aforementioned filters, imposes challenges to the SFH inferred solely by SED-fitting, regardless of the model complexity. We hence demonstrate for the first time that including the $B/R$ ratio itself as a constraint on top of the SED resolves such challenges and offers an alternative solution of SFH that can explain both the $B/R$ and SED for relevant galaxies. The solution constrained by also the $B/R$ ratio also has a lower Bayesian Information Criterion, compared with the solution constrained by the SED only. With the higher spatial resolution of the $B/R$ ratio measurement and the higher sensitivity towards slightly older star formation history up to $\sim 50\,$Myr ago, the $B/R$ ratio could be a powerful method to study spatially resolved stellar population in relevant galaxies.

Our solution of Warhol's SFH after incorporating the $B/R$ ratio as a constraint on top of SED increases the estimated stellar mass formed in the last $\sim50\,$Myr by a factor of $\sim 60$ for the whole Warhol arc, and a factor of ten for region 2, a subset of Warhol, with $3\sigma$ confidence. Although this change is subtle as it nevertheless only contributes to a small fraction of the stellar mass of the entire Warhol; this kind of study could drastically change our understanding of galaxies in the early universe, when we are witnessing the growth of their stellar mass \citep[e.g., ][]{Weaver_2023, Weibel_2024}. Also, the current signal-to-noise in the $B/R$ ratio is low and has inferred a low significance in the change in inferred stellar mass. More transient observations will further test the significance of the $B/R$ ratio as a constraint on SFH.

The current record holder of lensed star candidates has a redshift up to $\sim 6$ \citep{Earandel}. With sufficient observation, thus transient detection at galaxies across the cosmic history, the $B/R$ ratio can be a powerful tool in better assessing the star formation of galaxies in the universe as early as $\lesssim 1\,$Gyr, and therefore the understanding of the evolution of galaxies over cosmic time. Another interesting direction of investigation along the line is whether this change in stellar mass formed over the last $\sim 50\,$Myr in relevant galaxies is a systematic bias as revealed by the $B/R$ ratio. Highly resolved lensed galaxies with many transients detected by both {\it HST} and {\it JWST}, such as the ``Dragon'' \citep[$z = 0.73$ with $40+$ transients][]{Kelly_2022_Flashlights, Fudamoto_2025}, would serve as an excellent test.

\begin{acknowledgements}
    
We sincerely thank the referee for the comments, which have eventually led to a more comprehensive analysis. We also thank Yoshinobu Fudamoto, Juno Li, and Arsen Levitskiy for the insightful discussion. 

S.K.L., J.L., and J.N. acknowledge support from the Research Grants Council (RGC) of Hong Kong through the General Research Fund (GRF) 17302023. 
J.M.P. received financial support from the Formación de Personal Investigador (FPI) programme, ref. PRE2020-096261, associated with the Spanish Agencia Estatal de Investigaci\'on project MDM-2017-0765-20-2.
J.M.D. acknowledges support from project PID2022-138896NB-C51 (MCIU/AEI/MINECO/FEDER, UE) Ministerio de Ciencia, Investigaci\'on y Universidades.
J.N. also acknowledges the support of the Dissertation Year Fellowship issued by the University of Hong Kong.
P.L.K. and L.L.R.W. acknowledge the support by NASA/HST grants GO-15936 and GO-16278 from STScI, which is operated by the Association of Universities for Research in Astronomy, Inc., under NASA contract NAS5-26555.
AZ acknowledges support by grant 2020750 from the United States-Israel Binational Science Foundation (BSF) and grant 2109066 from the United States National Science Foundation (NSF), and by the Israel Science Foundation Grant No. 864/23.

We made use of the following software: Python, NumPy, Matplotlib, SciPy, Astropy, {\tt SPISEA}, emcee, and corner.
\end{acknowledgements}

\bibliography{sample7}{}
\bibliographystyle{aasjournal}

\appendix
\onecolumn
\section{Selected simulated cases}

\vspace{1em}
\begin{figure*}[h!]
\centering
    \includegraphics[width=\linewidth]{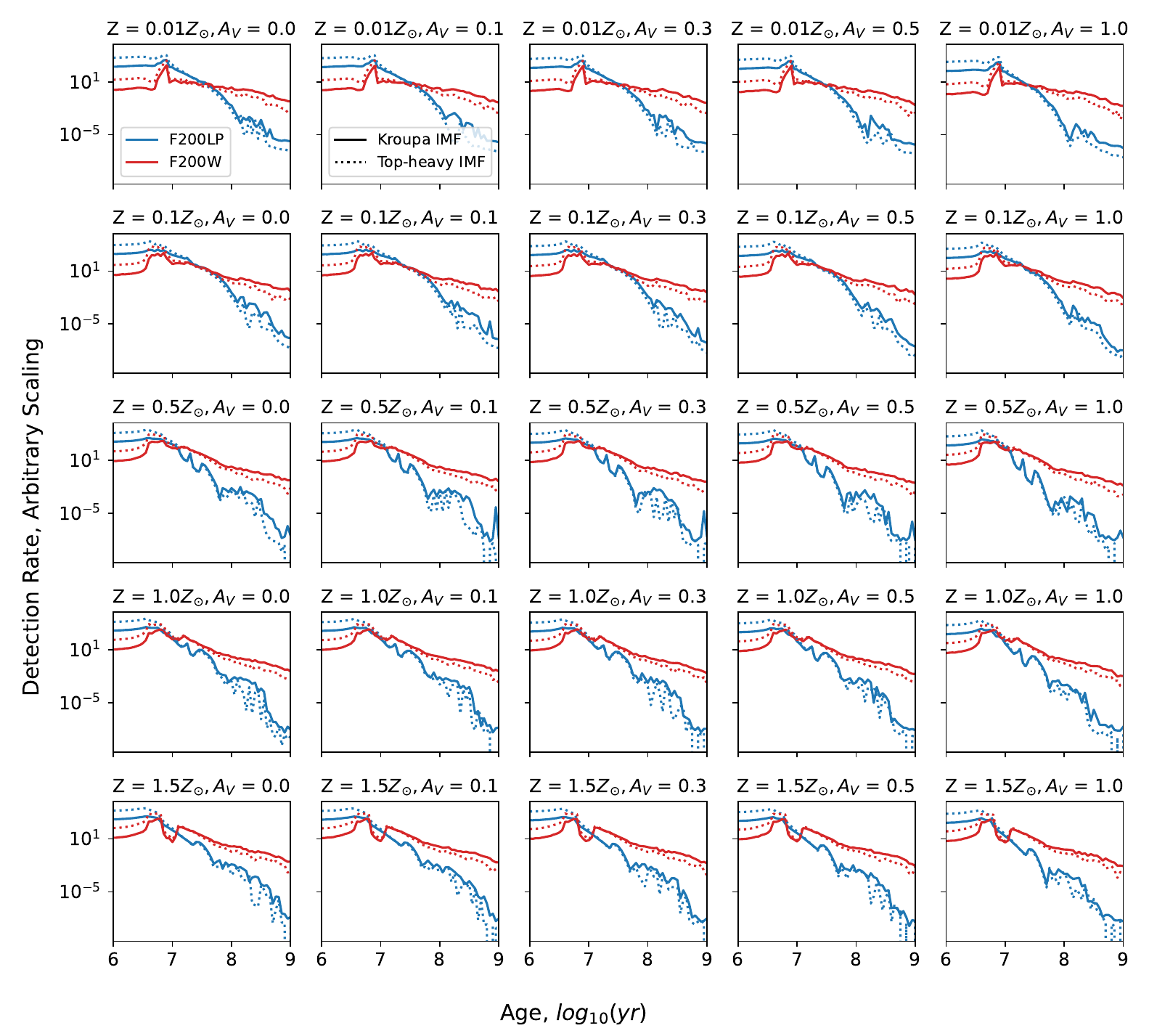}
    \caption{Same as Fig.~\ref{fig: detection_rate}, but for the cases with different metallicities and dust extinction as denoted in the subtitles of each of the panels. We do not show the error bar for clarity issues.}
    \label{fig: detection_rate_all}
\end{figure*}

\newpage

\section{MCMC analysis}
\label{sec: mcmc}

We use emcee to carry out our MCMC analysis throughout this paper. The model likelihood, $L$, is given as:

\begin{equation}
    L \propto \textrm{exp}[-\frac{1}{2} \sum ^{N}_{i}(\frac{x_{i}(\theta) - \mu_{i}}{\sigma_{i}})^{2}] \, ,
\end{equation}

\noindent where $x_{i}(\theta)$ is the model prediction given a set of parameters to be optimized upon, $\theta$, whereas $\mu_{i}$ is the observed value with uncertainty $\sigma_{i}$ over $N$ data points, $i$.

\vspace{1em}
\begin{figure}[h!]
\centering
    \includegraphics[width=0.85\linewidth]{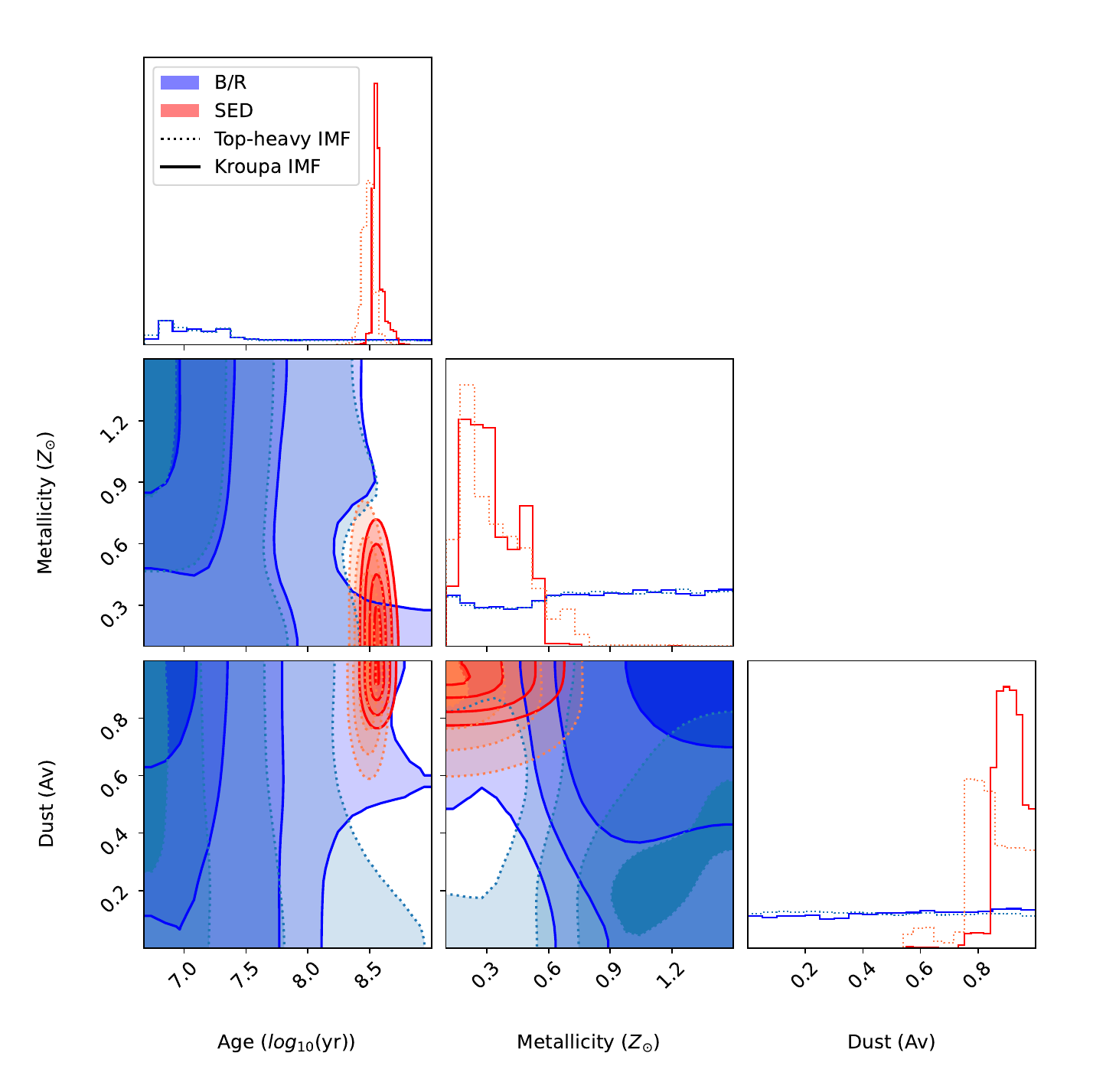}
    \caption{Markov-Chain Monte-Carlo analysis regarding the constraining power of $B/R$ (blue) and SED (red) of Warhol on the parameter space, given that we assume there is only one single burst of star formation at age $t$ with metallicity $Z_{\star}$ and dust extinction $A_{V}$. Since we only care about the inference of these three parameters, all the SED templates generated by SPISEA are pre-normalized to best fit the observed SED of Warhol. The contour levels correspond to $1-4\sigma$ level confidences. Histograms/contours in solid lines represent the case for a Kroupa IMF, where those in dotted lines (with a slight color tweak to improve visibility) represent the case for a Top-heavy IMF. There is no significant difference in preference for star formation parameters constrained by either $B/R$ or SED when a different IMF is adopted.}
    \label{fig: mcmc}
\end{figure}

\vspace{1em}
\begin{figure*}[h!]
\centering
    \includegraphics[width=.9\linewidth]{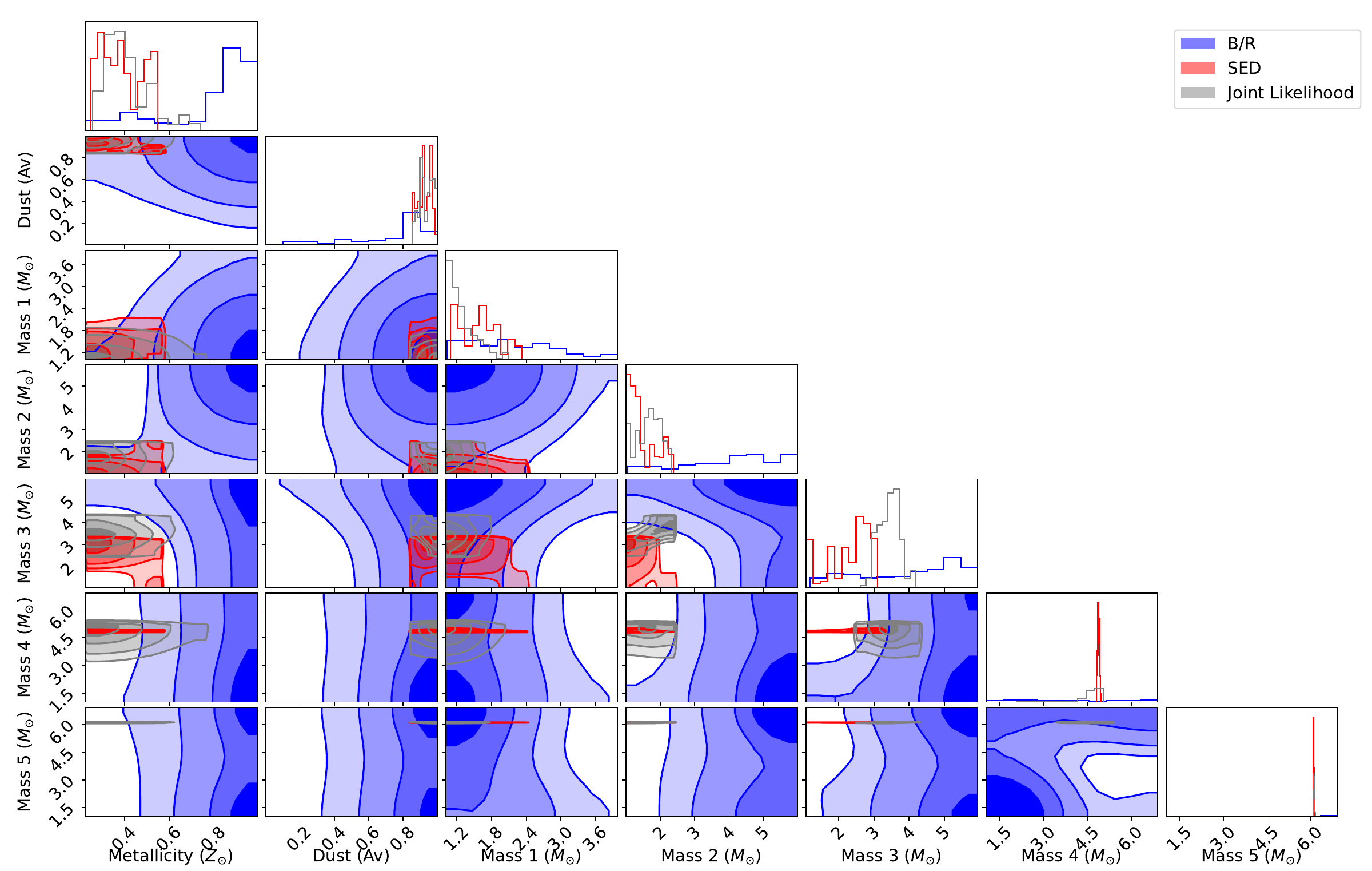}
    \caption{Same as Fig.~\ref{fig: mcmc}, but with a non-parametric SFH instead of two single star bursts. Since we concluded that the IMF is not important for our analysis, here we only show the case where a Kroupa IMF is adopted. The gray contour represents the case where both $B/R$ and SED are used as constraints. Mass 1-5 corresponds to the mass formed in the 5 age bins of $1-5\,$Myr, $5-10\,$Myr, $10-50\,$Myr, $50-250\,$Myr, and $250\,$Myr$-1\,$Gyr, respectively. The MCMC here corresponds to the results shown in Fig.~\ref{fig: SED_fit_nonparam} and Fig.~\ref{fig: SFH_nonparam}.}
    \label{fig: mcmc_nonparam}
\end{figure*}

\vspace{1em}
\begin{figure*}[h!]
\centering
    \includegraphics[width=.9\linewidth]{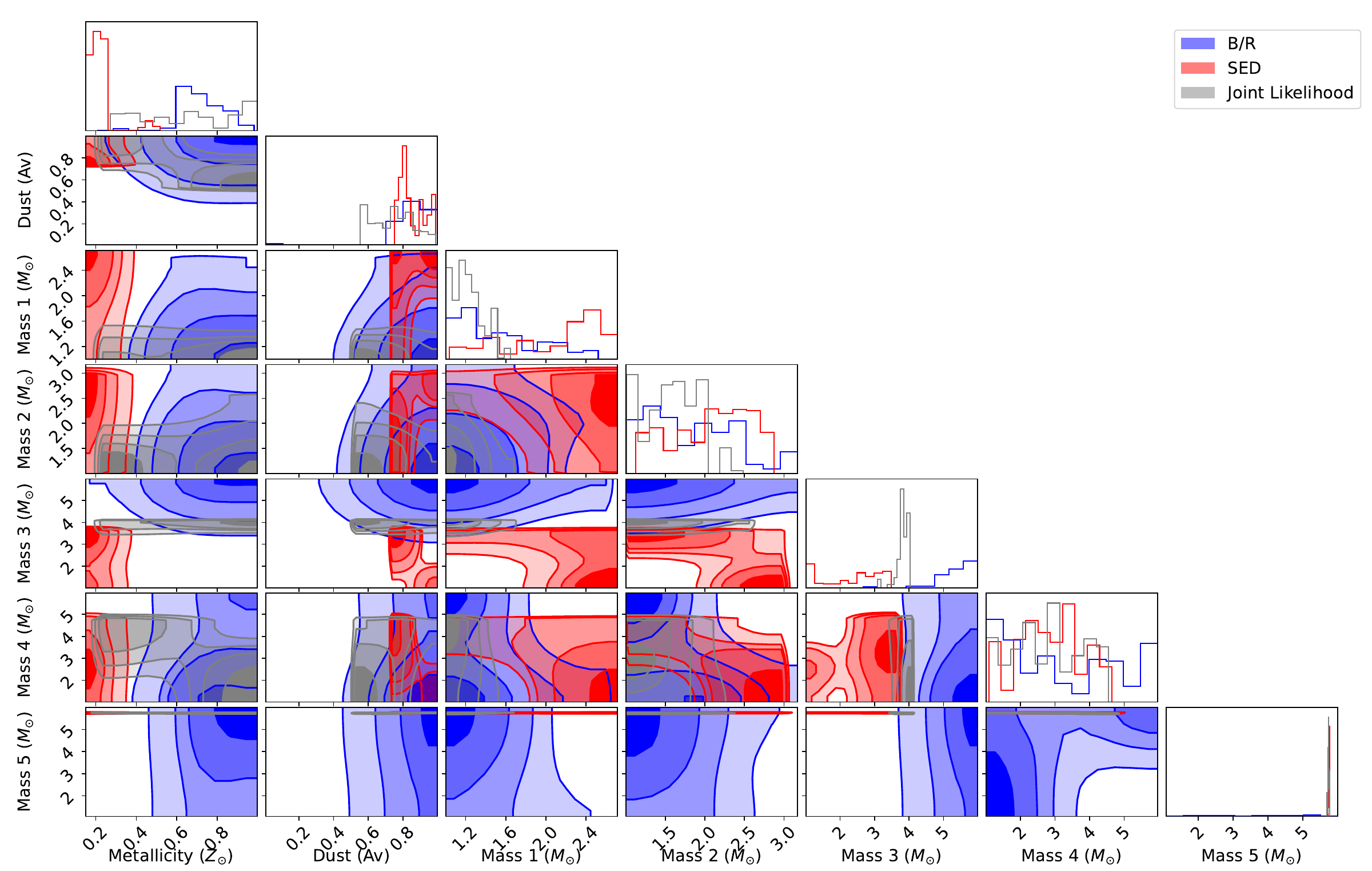}
    \caption{Same as Fig.~\ref{fig: mcmc_nonparam}, but confined the analysis to region 2 of Warhol. The MCMC here corresponds to the results shown in Fig.~\ref{fig: Region2_SED} and Fig.~\ref{fig: Region2_SFH}.    }
    \label{fig: mcmc_nonparam_R2}
\end{figure*}

\end{document}